%% file: manuscript.tex
\documentclass[a4paper,fleqn,useAMS,usenatbib]{mnras}

\pdfoutput=1

\usepackage{mathptmx}

\usepackage[T1]{fontenc}
\usepackage{ae,aecompl}
\usepackage{subfigure}
\usepackage{stackengine}

\usepackage[dvips]{graphicx}
\usepackage{amsmath}
\usepackage{amsfonts}
\usepackage{amssymb}
\usepackage{comment}
\usepackage{pdflscape}
\usepackage[dvipsnames]{xcolor}
\usepackage[%
        ]{hyperref}
        
\hypersetup{
	colorlinks=true,	
	urlcolor=MidnightBlue,		
	pdfpagelabels=true,
	hypertexnames=true,
	plainpages=false,
	naturalnames=true,
	pdftitle={WD transiting planets with LSST},    
	pdfauthor={Cort\'{e}s \& Kipping},     
	linkcolor=WildStrawberry,          
	citecolor=ForestGreen,        
}

\input{shortcuts.tex}

\title[WD transiting planets with LSST]{On the detectability of transiting planets orbiting white dwarfs using LSST}
\author[Cort\'{e}s \& Kipping]{Jorge Cort\'{e}s$^{1}$\thanks{E-mail:
\href{mailto:jorgecortes@astro.columbia.edu}{jorgecortes@astro.columbia.edu}} \& David Kipping$^{1}$\\
$^{1}$Dept. of Astronomy, Columbia University, 550 W 120th Street, New York NY 10027
}

\date{Accepted . Received ; in original form }

\pubyear{2018}

\begin{document}
\label{firstpage}
\pagerange{\pageref{firstpage}--\pageref{lastpage}}
\maketitle

\begin{abstract}
White dwarfs are one of the few types of stellar objects for which we know almost nothing about the possible existence of companion planets. Recent evidence for metal contaminated atmospheres, circumstellar debris disks and transiting planetary debris all indicate that planets may be likely. However, white dwarf transit surveys are challenging due to the intrinsic faintness of such objects, the short timescale of the transits and the low transit probabilities due to their compact radii.
The Large Synoptic Survey Telescope (LSST) offers a remedy to these problems as a deep, half-sky survey with fast exposures encompassing approximately 10 million white dwarfs with $r<24.5$ apparent magnitude. We simulate LSST photometric observations of 3.5 million white dwarfs over a ten-year period and calculate the detectability of companion planets with $P<10$\,d via transits. We find typical detection rates in the range of $5 \times 10^{-6}$ to $4 \times 10^{-4}$ for Ceres-sized bodies to Earth-sized worlds, yielding $\sim 50$ to $4\,000$ detections for a 100\% occurrence rate of each. For terrestrial planets in the continuously habitable zone, we find detection rates of $\sim 10^{-3}$ indicating that LSST would reveal hundreds of such worlds for occurrence rates in the range of 1\% to 10\%.
%
\end{abstract}


\section{Introduction}
\label{sec:intro}

\input{introduction.tex}

\section{Simulating LSST Light Curves}
\label{sec:simulations}

\input{simulations.tex}

\section{Planet Injection \& Recovery}
\label{sec:injections}

\input{injections.tex}

\section{Discussion}
\label{sec:discussion}

\input{discussion.tex}

\section*{Acknowledgments}

DMK is supported by the Alfred P. Sloan Foundation. JC gratefully acknowledges the support of the Columbia University Bridge to Ph.D. Program in the Natural Sciences and the support provided by the NSF through grant AST-1539931. 
Special thanks to Scott Daniel for his guidance with using \catsim\ and to Jay Farihi for his helpful comments on this work.
Thanks to the members of the Cool Worlds Lab for regular discussions and
feedback during the preparation of this work.

%

\bsp
\label{lastpage}
\end{document}

%% file: shortcuts.tex
\newcommand{\kepler}{{\it{Kepler}}}
\newcommand{\opsim}{{\tt OpSim}}
\newcommand{\catsim}{{\tt CatSim}}
\newcommand{\logg}{\log g}
\newcommand{\teff}{T_{\mathrm{eff}}}
\newcommand{\snr}{\mathrm{S/N}}
\newcommand{\batman}{{\tt batman}}
\newcommand{\fatboy}{{\tt fatboy}}
\newcommand{\forecaster}{{\tt forecaster}}

\newcommand{\actuallink}{https://github.com/CoolWorlds/whiteworlds}


%% file: introduction.tex
The vast majority of stars end their lives as white dwarfs (WDs). Within these
stellar remnants, nuclear fusion has ceased and thus further inward collapse
is resisted by electron degeneracy pressure. As a result, these
remnants have compact radii similar to that of the Earth, yet with a mass of
typically half of that of the Sun. Despite the absence of internal fusion,
these stars shine for billions of years as they slowly cool, providing a means
of studying their behavior and environment.

In recent years, there has been increasing interest in searching for and
studying planets orbiting ever smaller primaries \citep{reich:2013}. In the
early years of the modern exoplanet era, surveys typically focused on FGK
stars resembling the Sun (e.g. \citealt{bakos:2004,pollacco:2006,bakos:2013,
wilson:2008}). A basic argument was that such stars clearly have a credible
chance of supporting life as demonstrated by our own existence, but smaller
primaries (in particular M-dwarfs) may be less favorable\footnote{We
note that more recent studies take a more optimstic view of M-dwarf
habitability; for example see the review of \citealt{shields:2016}.}
(e.g. see \citealt{dole:1964,kasting:1993}). Both radial velocity surveys,
such as HARPS \citep{bonfils:2013}, and photometric surveys, such as MEarth
\citep{charbonneau:2009,irwin:2015}, began to shift the focus towards smaller
M-dwarfs, arguing that their smaller dimensions provide a significant boost to
our sensitivity. The discovery by \kepler\ that early-type M-dwarfs appear to
host more planets, including habitable-zone planets, than Sun-like stars
\citep{dressing:2015} has brought M-dwarfs keenly into focus of planet hunters
in recent years.

Surveys such as SPECULOOS \citep{burdanov:2017} aim to push down further to
late-type M-dwarfs as suitable targets for hunting planets. Surveys for planets
orbiting even smaller and fainter brown dwarfs appear imminent
\citep{triaud:2013}. Clearly then, the field of exoplanets has departed from
the paradigm that we should only survey types of stars where we know for
certain life is possible. There are certainly many challenges to life as we
know it surviving and thriving on planets such as Proxima b
\citep{anglada:2016} with a rich and active debate taking place in the
literature \citep{ribas:2016,turbet:2016,garcia:2017}. Against this backdrop,
an open-minded philosophy for allocating observational resources and effort is
to focus on looking for life in places where we have the ability to
observationally test it, not necessarily the places which we hypothesize as
being the most habitable.

It could be argued that the pinnacle of this drive towards ever smaller planet
hosts is represented by white dwarfs\footnote{since more compact objects are
not luminous}. With a radius of about ten times less than that of the smallest
M-dwarf \citep{shipman:1979,chen:2017}, these stars provide a major
amplification of transit signals of up to 100-fold. On the downside, these
stars are intrinsically faint, meaning that they are quite uncommon in a
magnitude-limited survey. Further, their small dimensions mean that transits
would last for minutes, not hours, posing a challenge to conventional surveys
whose integration times are often too long to resolve the signals
\citep{jenkins:2010}.

The potential value of a WD exoplanet survey was highlighted by
\citet{agol:2011}, who argued that at least a few thousand WDs need to be
surveyed to place meaningful constraints on their existence. The largest survey
of WDs for transiting planets to date was recently published by
\citet{sluijs:2017}, who found no examples amongst $1\,148$ WDs observed by \textit{K2}.
If such planets could be found, \citet{loeb:2013} estimate that
only a few hours of JWST time would be needed to detect biosignatures in their
atmospheres (should they exist) thanks to the small size of the host.

At first, it may seem a stretch to consider WDs as potential planet hosts.
These stars must have passed through the red giant phase engulfing any planets
within an AU \citep{sandquist:1998,sandquist:2002}. Despite this, there are now
numerous indirect clues suggesting planets may indeed orbit WDs. First,
$\sim 30$\% of WDs appear to have metal contaminated atmospheres, indicative of
a continuous supply of in-falling rocky material as a result of the rapid
convection times \citep{zuckerman:2003,zuckerman:2010,koester:2014}. Second,
debris disks appear common around WDs, with a lower limit being that 5\% host
such disks potentially supporting a second generation of planet formation
\citep{barber:2012}. Third, there is direct evidence of a likely disintegrating planet
orbiting WD 1145+017 \citep{vanderburg:2015}. Put together, these clues
strongly motivate that we should at least attempt a deep search for planets
orbiting white dwarfs.

As alluded to earlier though, there are two significant obstacles facing
any survey attempting to seek WD planets. First, WDs are faint and
thus in order to survey a large number we need to survey both a large fraction
of the sky and go deep \citep{kilic:2013}. Second, the transits last for
minutes, meaning that exposures must not exceed that timescale in order to
avoid significant distortion and dilution of the transit morphology
\citep{binning:2010}. LSST is essentially completely unique in being able to
overcome these two challenges. The sample size issue will be certainly
overcome, since LSST will observe down to $24^{\mathrm{th}}$ magnitude,
including an expected $\sim 10^7$ WDs \citep{agol:2011}. The problem of
transit distortion is also overcome since LSST is expected to take two 15
second exposures back to back in normal operation, sufficient to avoid smearing
of the light curve \citep{abell:2009}.

The most obvious drawback in using LSST for this purpose is that it does not
survey each patch of the sky for long periods of time. Thus, one should never
reasonably expect a particular instance of WD transit to be covered, only a
partial transit (with the exception of the deep drilling fields). However, if
the signal is strictly periodic, then it is not the temporal coverage which we
actually care about but rather the phase coverage \citep{lund:2015}. Over the ten year baseline of LSST we should expect well-suited phase coverage for any given star\footnote{This has been previously demonstrated to certainly be true for transiting planets of normal stars, see \citet{jacklin:2015,jacklin:2017}.}.

For the reasons described above, we hypothesized that LSST would be an excellent machine 
for searching for WD transiting planets. However, a detailed study injecting real transit signals into
LSST-like sparsely sampled time series around realistic WDs is notably
absent in the literature and thus reasonable concerns might exist about the
true feasibility of detecting WD planets with LSST. In response to this, the
work presented in what follows offers a detailed suite of simulations of
planets injected into LSST light curves to evaluate their detectability.

%% file: simulations.tex
\subsection{Overview: A Monte Carlo Approach}
\label{sub:simsoverview}

The primary objective of this work is to evaluate the detectability of
transiting planets with LSST. An unconditional yield estimate is not formally
calculable in the absence of any information about the occurrence rate and
distribution of planets orbiting WDs. Instead, we aim to ask the question, if
a WD hosted a planet with radius $R_P$ and orbital period $P$, how detectable
would that planet be using LSST\footnote{We return to the issue of yield
estimation later in Section~\ref{sec:injections}, where we show that a yield
estimate can be estimated conditioned upon specific assumptions about the
underlying distribution of WD planets.}?

This question is tackled from a Monte Carlo perspective using numerical
simulations. Whilst it may certainly be possible to express a reasonable
parametric model describing the detectability of WD planets using analytic
arguments, it is clearly complicated by the sparse non-uniform scheduling
expected with LSST \citep{abell:2009}. A Monte Carlo approach is attractive
if there exists a means of generating representative photometric time series
expected from LSST for WDs. Unpacking that statement, the requirements of
such an approach can be more specifically stated as being a) the need
to simulate a representative distribution of the properties of WDs that will
be observed by LSST b) the need to simulate representative photometric
time series expected of said stars, accounting for realistic LSST noise and
scheduling constraints.

Fortunately, both of these requirements are satisfied by software resources
made available by the LSST team, namely \opsim\ \citep{coffey:2006} and
\catsim. Using these tools then, we describe in what follows how we generate
representative photometric time series of WDs that will be observed by LSST.
We discuss our approach for measuring planet detectability later in
Section~\ref{sec:injections}.

\subsection{Generating a WD Catalog with \catsim}
\label{sub:catsim}

LSST's Catalog Simulator (\catsim) is used to incorporate a realistic distribution 
of WDs within the Milky Way Galaxy. WDs, and all stars accessed by \catsim\, 
are generated by \textit{galfast}, a GPU-accelerated package that fits models of a 
thick and thin disk, and halo to SDSS data \citep{juric:2008}. The default simulated
universe accessed by the \catsim\ stack is stored as a database on \fatboy, a machine located at
the University of Washington. Information for all WD's with $\teff \leq 11\,000$K
was retrieved and stored locally. The temperature constraint is imposed in consideration of 
the white dwarf habitable zone \citep{agol:2011}. 


\subsection{Interpolating Stellar Properties}
\label{sub:starprops}

Stars listed in the \fatboy\ database state the $\logg$ and $\teff$ of each source. 
However, in order to inject a planet around a given WD, we need to know the stellar
mass, $M_{\star}$, allowing us to convert a chosen orbital period into
semi-major axis via Kepler's Third Law\footnote{We note that a stellar density
would also suffice, but this is also unavailable.}. Rather than simply adopt a
uniform stellar density for all WDs, we seek to create the most realistic
catalog possible in this work. Accordingly, we elected to estimate a realistic
stellar mass for each WD based on the provided \catsim\ information.

To accomplish this, we used the evolutionary cooling models of hydrogen-and helium-atmosphere white dwarfs\footnote{Available at
\href{http://www.astro.umontreal.ca/~bergeron/CoolingModels}{this URL}} from
\citet{holberg:2006,kowalski:2006,tremblay:2011,bergeron:2011}. We perform a bi-linear interpolation of these model grids such that for any combination of
$\logg$ and $\teff$, we are able to assign a unique stellar mass. We find that
both the H-rich \& He-rich WDs masses approximately follow a normal distribution peaked
at half a solar mass with a $\sim 0.1$\,$M_{\odot}$ standard deviation.

\subsection{Generating Light Curves with \opsim\ and \catsim}
\label{sub:opsim}

To generate mock observations, LSST's Operations Simulator (\opsim) is used 
in conjunction with \catsim \footnote{Tutorial notebooks from LSST available at \href{https://github.com/uwssg/LSST-Tutorials/tree/master/CatSim}{this URL}.}. 
\opsim\ provides a sample cadence for LSST's ten-year 
observing strategy; we make use of the current reference run, 
\textit{minion\textunderscore1016}. As it currently stands, the \catsim\ stack 
will only produce light curves for objects that incorporate a variability model; thus, we set the variability constraint within {\tt LightCurveGenerator.py} of \catsim\ to \lq None\rq. At this point, we are able to retrieve light curve data (time, magnitude, errors) for all WDs. 

\begin{figure*}
\begin{center}
\includegraphics[width=15cm, angle=0]{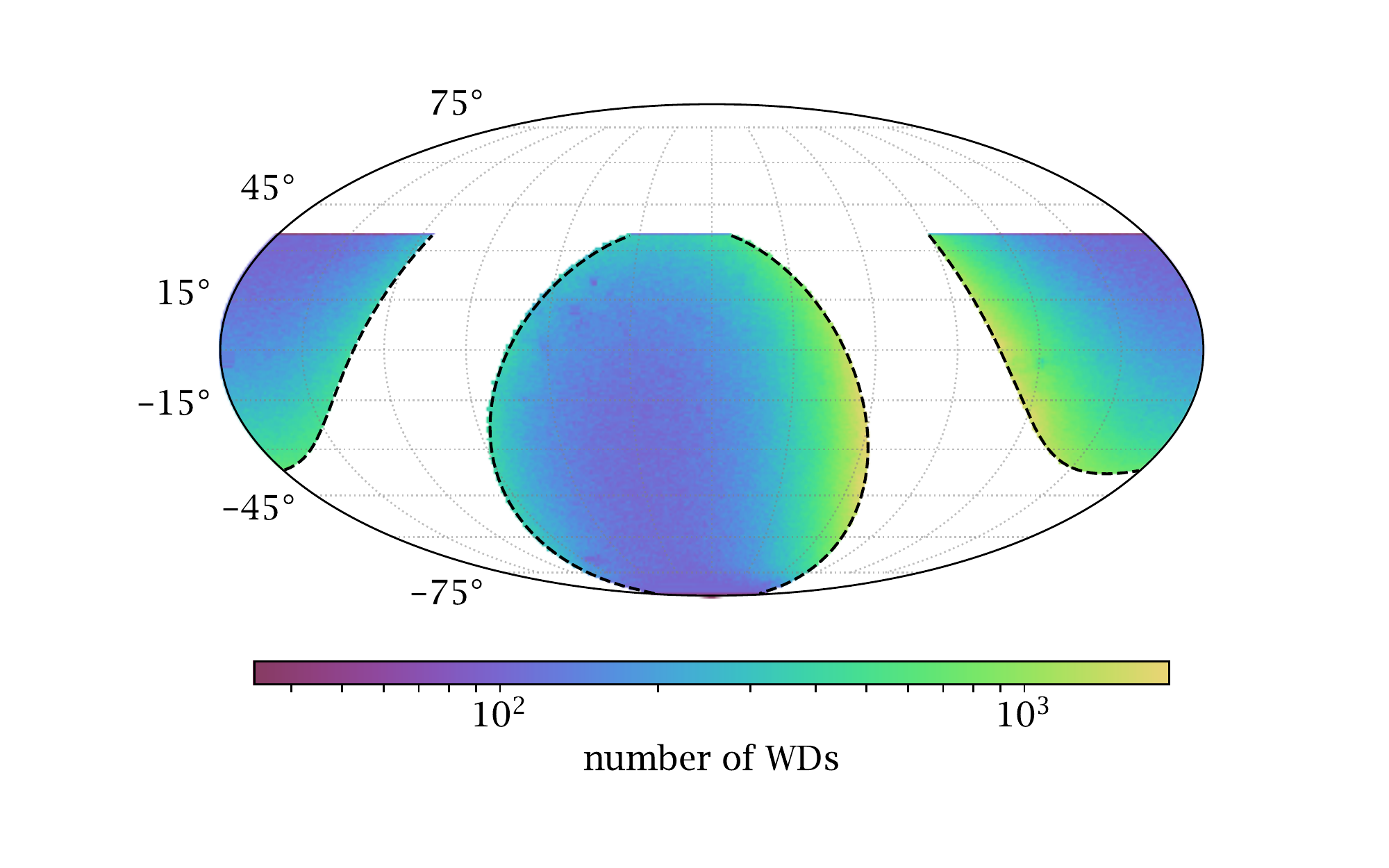}
\caption{Number density of \catsim\ generated WDs used in our survey. The galactic
plane is masked to within $25^{\circ}$ in this work.}
\label{fig:distribution}
\end{center}
\end{figure*}

To avoid memory issues on our local machine, we excluded data for WDs within $\pm25^\circ$ of galactic longitude, as seen in Figure~\ref{fig:distribution}. Additionally, this constraint has the added benefit of exploring regions which are less affected by crowding\footnote{Whilst crowding is not an issue for simulations, such as those used here, crowded fields may lead to source contamination during the actual LSST observations.}. All WD light curves were stored locally on a database file. 

We briefly point out that the light curves generated assume that the visit-to-visit 
calibration error is much less than that of the point-to-point photometric 
uncertainty. Given our sources are generally faint, the point-to-point 
uncertainties are already large meaning that if LSST achieves visit-to-visit 
calibrations of order of a percent or better, this is unlikely to be a meaningful 
source of error\footnote{Unlike the case of Sun-like stars where likely sub-mmag 
long-term calibration would be needed}.

%% file: injections.tex
\subsection{Overview}

In order to calculate the detectability of planets around WDs with LSST,
it is of course necessary to inject planets into our synthetic light curves
described in Section~\ref{sec:simulations}, and so we turn our attention
to this here.

As touched on earlier, the distribution and occurrence rate of planets
around WDs is broadly unknown, making yield estimates, at best, conditional
(we define what we mean by conditional in detail shortly).
Accordingly, we place our emphasis here on estimating the marginalized
detectability of a planet of a given size and period. The process of
marginalization is discussed in Section~\ref{sub:sensitivity}.

\subsection{Injecting planets}
\label{sub:injection}

Using a random subset of 3.5 million of our $10^7$ simulated WD light curves, we inject a single planet around
every star with a random orbital period and physical radius. Periods are drawn
from a log-uniform distribution between 0.15\,d and 10\,d, and radii
from a log-uniform distribution from $\tfrac{1}{16}$\,$R_{\oplus}$ to
$16$\,$R_{\oplus}$. A random impact parameter, \textit{b}, between $0$ and
$1+(R_P/R_{\star})$ is assigned to each planet, such that a transit is
guaranteed. The time of inferior conjunction, $\phi$, is randomly assigned for each, and
every planet is assumed to follow a strictly circular Keplerian orbit.

Light curves for the injected planets are simulated using \batman\ \citep{kreidberg:2015} assuming a
15\,s integration time and uniform limb darkening. In all cases, we assume that
the eventual ten-year time series is available for the analysis. Further, we also
highlight that we assume only a single planet for each star and that 100\% of the stars have planets (it is straightforward to scale our results for arbitrary occurrence rates later).

For each star, we
calculate the planet's signal-to-noise ratio (S/N) and assign a yes/no binary
flag as to whether LSST is deemed to be sensitive to said planet (using the
method described in Section~\ref{sub:SNR}). Accordingly, amongst the 3.5
million stars, we can select those that have planets within a local size- and
period-bandwidth of a particular choice of $R_P$ and $P$, and then simply count
up what fraction of the stars had planets that LSST was sensitive to. This would
therefore represent the marginalized sensitivity. This exercise could be
repeated but instead using only a subset of the 3.5 million stars - for example
taking just the bright end. In this way, the marginalized sensitivity can be
computed using whatever marginalized sample one desires. 

\subsection{S/N and sensitivity}
\label{sub:SNR}

In this work, we inject planets but we do not blindly recover them. The
sensitivity of LSST to a given planet is computed by evaluating the S/N
instead. This choice was largely motivated by computational practicality.
Running a box least squares blind search \citep{kovacs:2002} on
3.5 million light curves would represent a major computational challenge.

We define S/N as follows. If the data can be assumed to approximately follow
a diagonal multivariate normal distribution, then one may define the goodness-of-fit
of a specific model using Pearson's chi squared. Accordingly, we compare the
$\chi^2$ of a null flat line model through the data versus that of the
planet model. We assume that any long-term variability has been filtered
(e.g. with a high-pass filter) and that short-term variability is much
lower amplitude than the transit signals injected (which often approach a
100\% eclipse depth). For our $\chi^2$ test, we use the exact parameters for the planet
model as that used for the injection, and thus this is why our approach
here is certainly not a blind recovery. However, the difference between
these two merit functions can be used to define S/N as follows: 

\begin{align}
\snr &\equiv \sqrt{\chi^2_{\mathrm{null}} - \chi^2_{\mathrm{planet}}}.
\end{align}

In order to assess whether LSST is deemed to be sensitive or non-sensitive
to a particular planet, we use a simple S/N threshold. Therefore, signals
with a S/N exceeding the threshold are always defined as LSST-sensitive,
and otherwise insensitive. This is a simplifying assumption since real
transit surveys do not have step functions sensitivity curves at a particular
threshold but rather S-like curves centered around a certain value
(e.g. see \citealt{christiansen:2016}). Nevertheless, certainly recoverability
does saturate to unity beyond a certain point and thus a threshold is not
an unreasonable approximation. Ultimately, the true curve will not be known
until real data becomes available. We adopt a S/N threshold of 7.1 in what
follows, as this was the same value initially adopted by the \kepler\ team
\citep{jenkins:2010b} and thus provides a standardized point of comparison.

\begin{figure*}
\begin{center}
\includegraphics[width=17.4cm, angle=0]{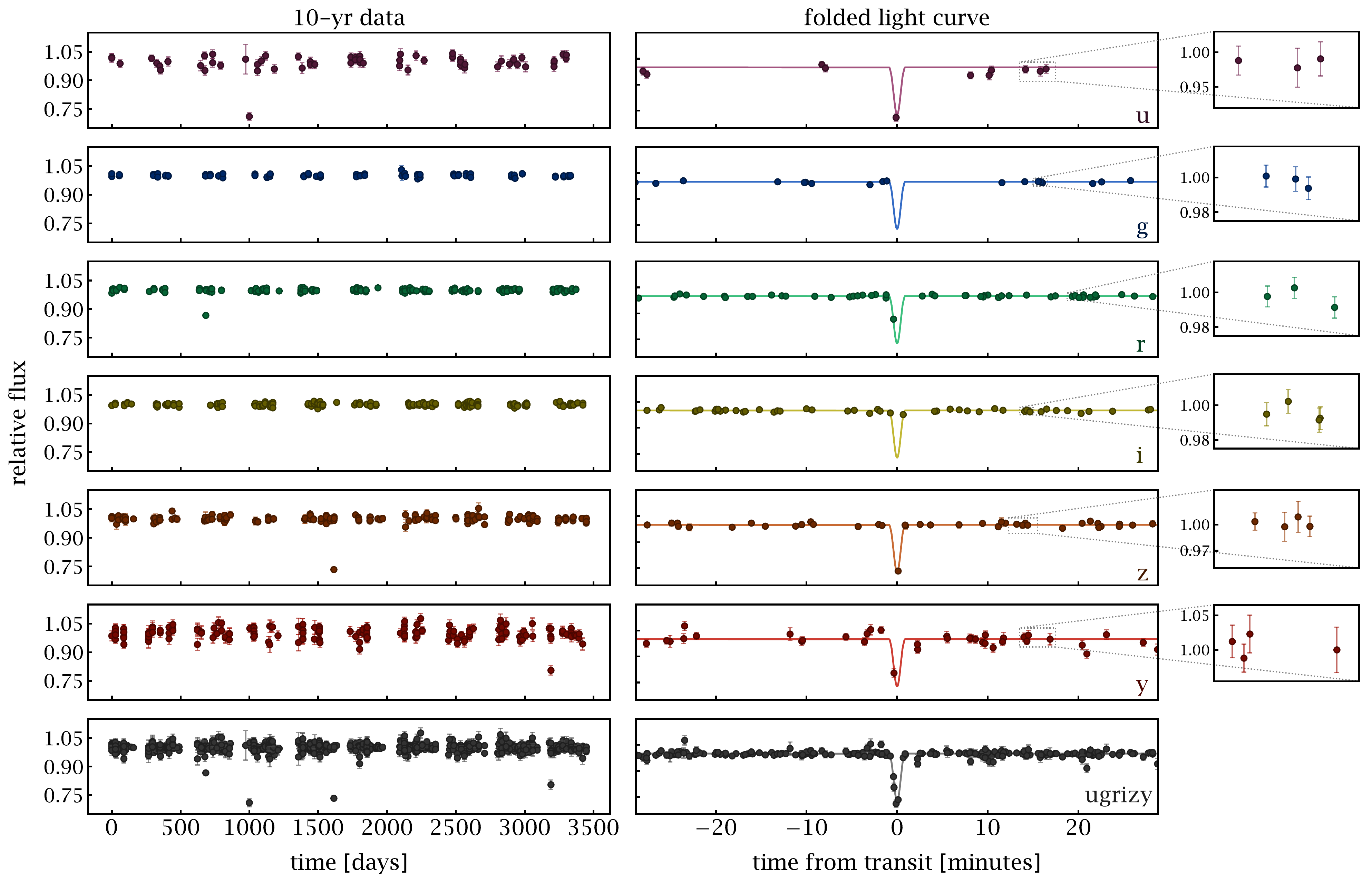}
\caption{Sample light curves demonstrating ten-year simulated observations (left) and phase folded data (right) across all six LSST filters (\textit{ugrizy}); injected planet light curve included for reference. Injected planet properties: $R_P/R_{\star} \approx{0.60}$, $P\approx{4.4\rm hr}$, $b\approx{0.63}$, $\phi\approx{1.4\pi}$}
\label{fig:sample_lightcurve}
\end{center}
\end{figure*}

In practice, the S/N is computed using all of the available LSST bandpasses
in conjunction to maximize our sensitivity to transit signals. An example of
this is shown in Figure~\ref{fig:sample_lightcurve} for illustration.

\subsection{Sensitivity in period-radius plane}
\label{sub:sensitivity}

After the calculations are complete, we have assigned a yes/no detection flag
to a total of 3.5 million individual simulated systems of randomized period
and radius. The sensitivity of LSST to each system is clearly quite varied
and depends on the star's magnitude, position, observing cadence, etc. To
simplify the picture, we calculate a so-called marginalized sensitivity
as a function of $R_P$ and $P$.

This is accomplished by first defining a local size- and period-bandwidth
and then moving across a two-dimensional grid of $R_P$ and $P$ and
calculating the local sensitivity as the the number of positives in that
window divided by the total stars in that window. In this way, the
estimate has marginalized (or averaged) over all other parameters, such as
stellar properties.

\begin{figure}
\includegraphics[width=\columnwidth]{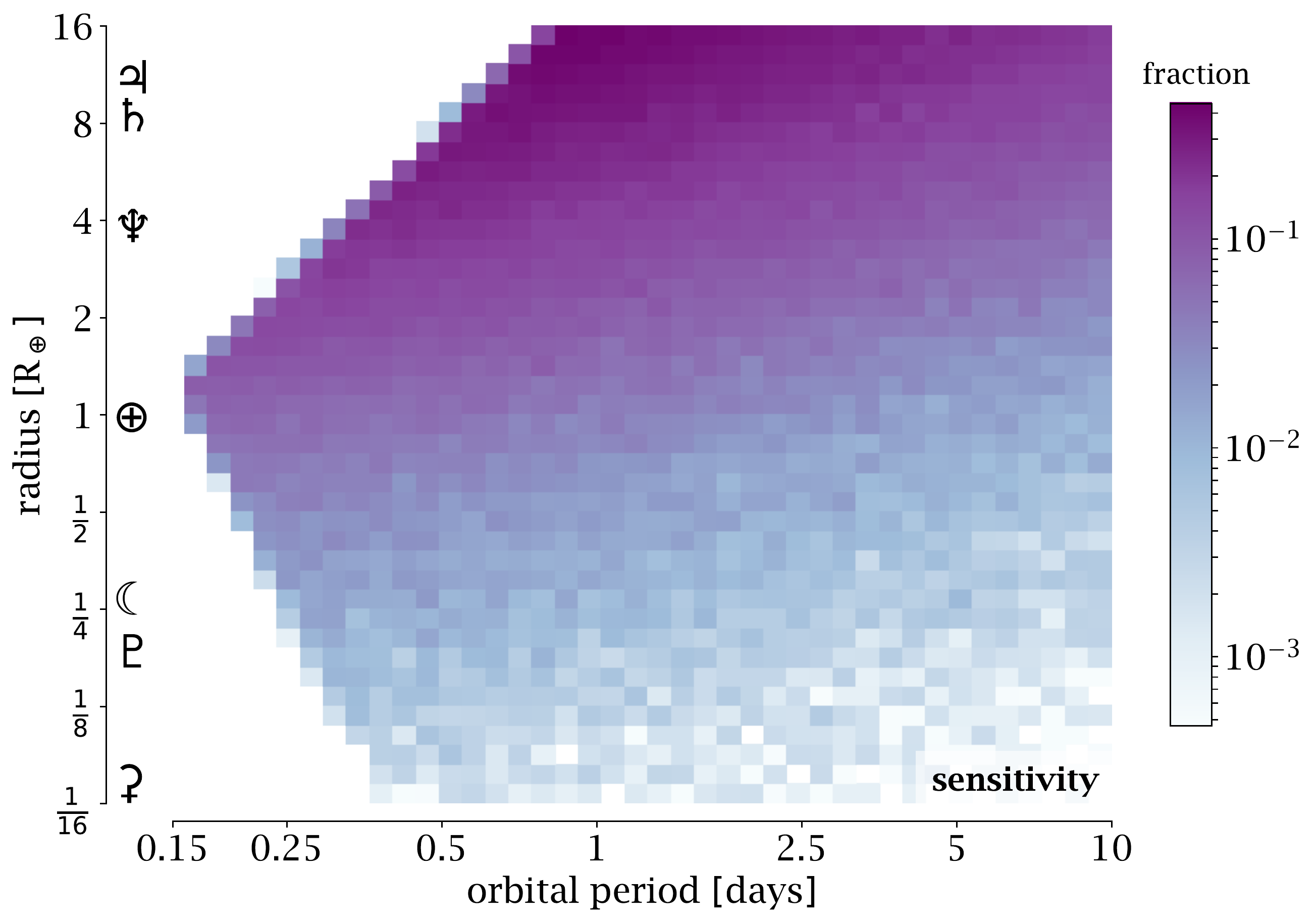}
\caption{Sensitivity results plotted as a function of period and radius from our LSST simulations.}
\label{fig:sensitivity}
\end{figure}

We define a bandwidth such that our periods and radii are divided into $40$ evenly spaced grid points in logarithmic space, which we found provides a good
balance of sufficient numbers per bin as well sufficient number of bins. The
resulting marginalized sensitivities are plotted in Figure~\ref{fig:sensitivity}
and are made publicly available at \href{https://github.com/jicortes/whiteworlds}{this URL}.
In our grid, it is apparent that we have removed some of the shortest period
objects from our sample. These censored objects fall within the Roche limit of
the star, where we have converted planetary radii into masses and then densities
using the \forecaster\ empirical mass-radius relation \citep{chen:2017}.


It is important to remember that sensitivity does not account for transit
probability and thus one might reasonably expect sensitivities approaching
unity for optimal cases. Indeed, the dynamic range apparent in Figure~\ref{fig:sensitivity}
reflects this. As expected, short-period planets are evidently more easily detected than their
longer period bretherin due to the increased frequency of their transits. The strongest bias occurs along the radius axis, where naturally larger planets are much more easily detected. 

Sensitivity has a dynamic range from zero to unity. It should be expected to saturate to unity in an exponential manner for highly idealized cases. Similarly, it should be expected to saturate to zero for extremely challenging cases. We therefore considered that sensitivity likely follows a logistic function. Accordingly, let us take a single slice along the log-period axis for a fixed choice of planetary size ($R_P=R_P'$). We would expect sensitivity along the log-period slice to be described by

\begin{align}
\mathcal{S} &= \frac{1}{1 - k \exp(\log (P/\mathrm{days}) - \log P_0)},
\label{eqn:Sscale}
\end{align}

where $k$ is a free parameter quantifying the steepness of the logistic curve and $\log P_0$ is a free parameter defining the mid-point.

We regressed this expression along all available choices of $R_P$ (removing cases where no detections were found), and found that $k$ is consistent across all choices with a mean and standard deviation of $-0.990 \pm 0.056$. In contrast, the $\log P_0$ term appears to linearly increase with respect to $R_P$. If we replace $\log P_0$ in Equation~\ref{eqn:Sscale} with a straight-line slope with respect to $\log R$, the result may be rearranged to the form,

\begin{align}
\mathcal{S} &= \frac{1}{1 + a (P/\mathrm{days})^{b} (R_P/R_{\oplus})^{c}},
\label{eqn:Sscale2}
\end{align}

where we find the values $a=18.77$, $b=0.393$ and $c=-0.943$ provide an excellent fit. The positive value for $b$ indicates that longer period planets are more difficult to detect, close to a $P^{-2/5}$ dependency. The reason why the scaling is better than $P^{-1/2}$, which one would expect if considering purely the transit frequency scaling, is due to the effect of increased durations at longer $P$ being preferentially detectable.

The negative $c$ coefficient indicates a roughly linear scaling of sensitivity with respect to planet size. The relationship is not quadratic, as one might naively expect, due to the fact that most of the detectable region of our parameter space occurs for $R_P>R_{\star}$, where quadratic scaling is not expected due to the total-eclipse and grazing-nature of the transits which dominate.

We highlight that directly comparing these coefficients to the theoretical expectations (e.g. from \citealt{sandford:2016}) is not generally possible since conventional transit yield/bias calculations do not operate in a regime dominated by grazing configurations. Nevertheless, Equation~\ref{eqn:Sscale2} with the quoted coefficients provides a straightforward way for the community to use our sensitivity results in other studies.

\subsection{From sensitivity to detectability}
\label{sub:completenessdef}

So far, our discussion has focused on sensitivity, which is defined
under the assumption that the impact parameter is uniformly distributed
between zero and $1+R_P/R_{\star}$. Thus, all stars are assumed to have
a transiting planet. Of course, even if all of the stars have planets,
they will not all host transiting planets. We therefore define
marginalized detectability as being similar to marginalized sensitivity
except we now account for the geometric transit probability expected
for each planet.

In the case of sensitivity, injected planets were either flagged as
detected (yes) or or non-detected (no). To compute detectability, we
multiply these essentially binary probabilities by the geometric probability of
$\mathrm{Pr}(0<b<1+R_P/R_{\star})$, which equals $(R_{\star}+R_P)/a$. Once each planet
has a detection probability assigned, we draw a random Bernouilli
integer to define the injected planet as being detected or non-detected
\footnote{We could have also used the probabilities themselves for the
subsequent marginalization, but we elected to binarize the results such
that they more closely resemble a real survey.}. We are then able to
compute a marginalized detectability in a similar fashion to that
described earlier for sensitivity.

\subsection{Detectability in period-radius plane}
\label{sub:completeness}

\begin{figure}
\includegraphics[width=\columnwidth]{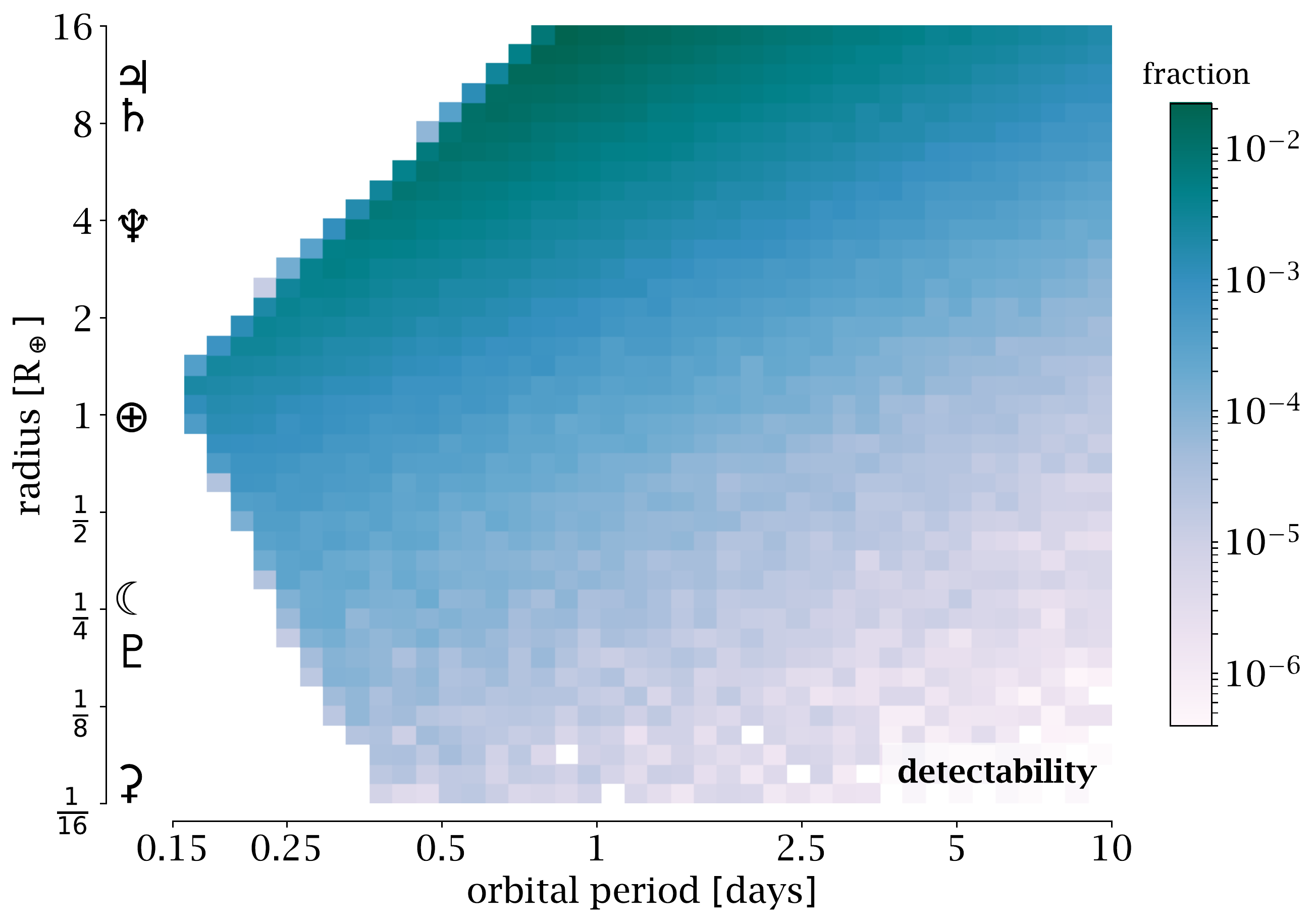}
\caption{Detectability, which accounts for the geometric transit probabilities,
plotted as a function of period and radius from our LSST simulations.}
\label{fig:detectability}
\end{figure}

Our detectability results, accounting now for transit probability, are
illustrated in Figure~\ref{fig:detectability} and are also made available
at \href{https://github.com/jicortes/whiteworlds}{this URL}. Broadly speaking, the
pattern appear similar to that of sensitivity, except the resulting scores
are typically two orders of magnitude lower, reflecting the $\sim 1$\%
transit probability of our injected planets. The dynamic range
increases due to longer period planets being particularly unlikely
to transit.

The mean detectability across all simulations equals 0.107\%,
implying that roughly $1$ in $1\,000$ WD systems harboring planets
need to be surveyed by LSST to bag a single detection.
We find that, marginalized across radius, the mean detectability
drops off as $\sim 1/P$, peaking at 1.5\% for the shortest Roche-stable
period possible ($\tfrac{1}{6}$ of a day) and dropping
down to 0.025\% at $P=10$\,d.

Using these numbers, it is straightforward to estimate yields
for various occurrence rates. If 100\% of the $\sim 10^7$ WD stars observed by LSST
harbor a planet with $P<10$\,d (i.e. $\eta=1$) then we would expect
$10\,700$ detections. Thus, we would require $\eta \lesssim 10^{-4}$ in
order for LSST to detect no examples of transiting WD planets. On this
basis, an LSST survey for WD transiting planets would be highly informative,
either delivering many examples of a new class of planetary system or
demonstrating to high confidence such systems are rare.

\begin{figure*}
\includegraphics[width=15cm]{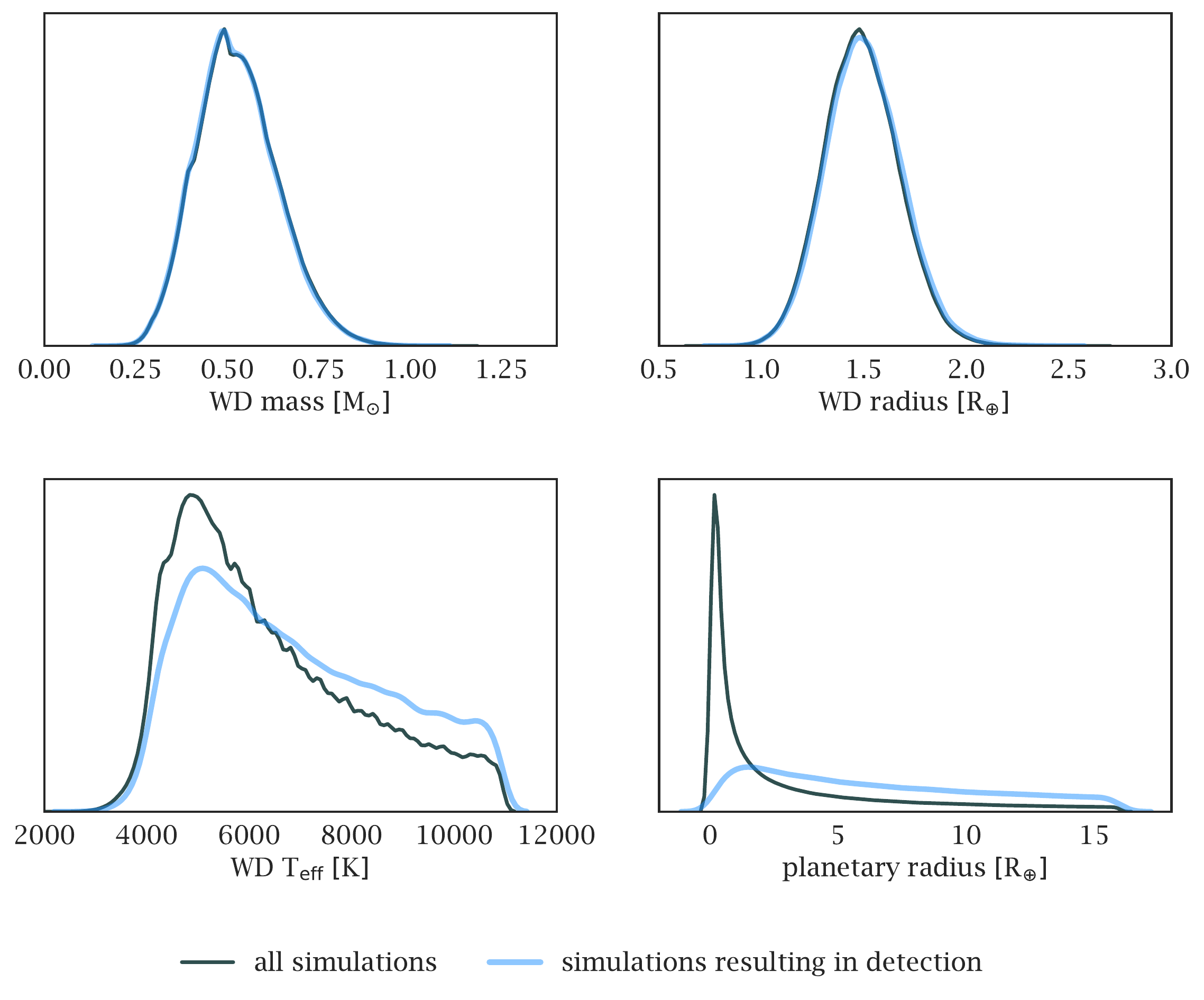}
\caption{Comparison of the system properties between the injected and detected populations.}
\label{fig:bias}
\end{figure*}

Although the log-uniform distributions in period and radius are likely as
good as a choice as any other, the more arbitrary choice in our prior
is the minimum/maximum period/radius. Changing these bounds will strongly
influence a marginalized yield estimate, such as that made above. Rather
than go through various hypothetical and speculative scenarios for these
bounds, we prefer to focus on the more robust kernel density estimates
for detectability (i.e. those shown in Figure~\ref{fig:detectability}),
which highlight that amongst a sample of $10^7$ WDs,
the vast majority of parameter space should be expected to yield
detections.

\subsection{Detection bias}
\label{sub:bias}

Armed with our numerical simulations, it is possible to investigate the impact of detection bias on our results. White dwarfs were generated as a representative astrophysical population using \catsim, but planet detections will clearly favor brighter stars and bigger planets. Figure~\ref{fig:bias} compares the detected vs injected population for four key parameters. We find that the stellar masses and radii of detected cases are representative of the true population. However, the detected stars tend to be hotter and thus more luminous making their photometric time series better quality for planet detections. The strong and expected bias towards larger planets is also recovered.


\subsection{Detectability of temperate transiting WD planets}
\label{sub:HZ}

A strong motivation behind the search for planets outside of our solar system is the potential characterization of habitable worlds, we thus turn our attention to a WD's temperate zone here. As discussed in \citet{agol:2011}, given a WD's mass and type (either H-rich or He-rich), a planet requires a specific orbital distance to fall within the continuously habitable zone (CHZ). This is defined as the range of orbital radii for which a planet will receive the necessary flux to sustain liquid water on its surface for at least 3 Gyr.  We take the outer limits of the CHZ for a H-rich and He-rich WD from \citet{agol:2011} and reassess the detectability results shown in Figure~\ref{fig:detectability}. We note that the inner limit of the CHZ comes up against the tidal disruption limit and thus can essentially be ignored in what follows.

\begin{figure}
\includegraphics[width=\columnwidth]{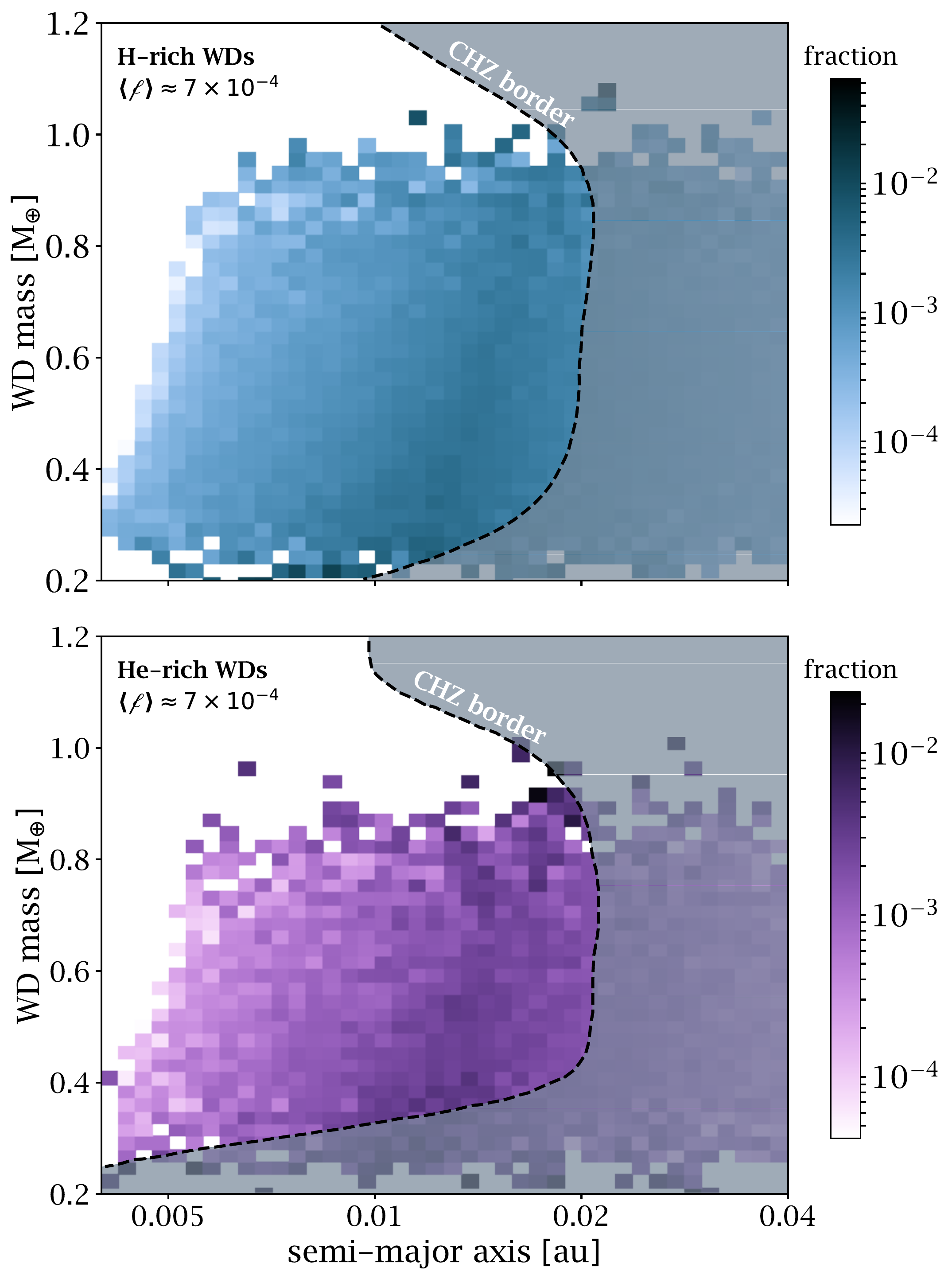}
\caption{Top: Detectability of planets around H-rich WDs with LSST via transits. Injected planets
are log-uniformly distributed in radius from 0.5 to 1.5\,$R_{\oplus}$ representing
terrestrial planets, many of which reside inside (to left of) the outer edge of the continuous habitable zone (CHZ) marked by the dashed line. Bottom: Same as top except for He-rich WDs. Average detection fraction within the CHZ is shown in the upper left for both.}
\label{fig:CHZ}
\end{figure}

We re-cast our detection figure in terms of WD mass and semi-major axis, so we can directly draw the CHZ contours from \citet{agol:2011} on top. In this way, the x-axis is essentially re-scaled via Kepler's Third Law from the version in Figure~\ref{fig:detectability}. The y-axis has replaced planetary radius with stellar mass, which was previously a marginalized quantity. It is therefore clear that in the revised version here, planetary radius will be a marginalized quantity. Specifically, we marginalize planets in the range $0.5 \leq (R_P/R_{\oplus}) \leq 1.5$ to focus on the potentially terrestrial-like planets. Finally, we split our sample into H-rich and He-rich stars, since the CHZ is noticeably different between the two \citep{agol:2011}. The final results are shown in Figure~\ref{fig:CHZ}.

We briefly remark that the detectability rates are in the range of $10^{-2}$ to $10^{-4}$, with the inner edge becoming attenuated due the impact of tidal disruption. This suggests that if the occurrence rate of CHZ rocky planets around WDs is $\eta \gtrsim 10^{-3}$, we should anticipate detections with LSST.

%% file: discussion.tex
We have demonstrated that LSST will have the capability to detect transiting planets around white dwarfs. Over an assumed ten year baseline, the sporadic sampling of LSST combines together to provide the excellent phase coverage needed for succesful transit detection (also see \citealt{lund:2015,jacklin:2015,jacklin:2017}). Further more, LSST's short integration time of 15 seconds ensures that WD transits, which typically last for a minute or two, are not significantly distorted due to binning \citep{binning:2010} aiding their identification. These two advantages enable detection but it is LSST's incredible depth, exceeding $25^{\mathrm{th}}$ magnitude, which make LSST potentially a revolution in the quest to detect WD planets since the survey will observe $\sim 10^7$ WDs.

We find that detection rates for $P<10$\,d planets range from $10^{-6}$ for long-period Ceres-sized bodies to $10^{-2}$ for short-period Jovians. Rates are naturally highly dependent upon what type of planet is under consideration but our suite of results are made publicly available at \href{https://github.com/jicortes/whiteworlds}{this URL}. Yield estimates are not directly possible due to our lack of information about the WD planet population. However, as an example we highlight that if $\eta=10$\% of WDs host a Mars-sized planet with $P<10$\,d, we should expect $\sim 100$ detections with LSST, and thus one might reasonable expect hundreds of discoveries.

If terrestrial planets reside in the continuous habitable zones of WDs with a frequency greater than $\eta \gtrsim 10^{-3}$, LSST should be expected to detect examples. Assuming again $\eta=10$\% would yield $\sim 10^2$ detections. We would expect the brightest planet hosting WD in such an example to be $r \sim 18$ - $22$ and thus may be suitable for atmospheric and even biosignature characterization with JWST \citep{loeb:2013}.

We highlight some limitations of our study. First, our work assumes that the transit detection pipeline acts as a perfect detector for $S/N>7.1$ and thus we did not execute blind recoveries of injected signals. Although typical search algorithms are found to have high efficiencies in this regime (e.g. see \citealt{christiansen:2016}), white dwarfs have not been surveyed in great detail before. Second, we assume long-term photometric behavior is filterable and that short-term variations have an amplitude less than the formal uncertainties. Given the faintness of our targets, we argue this is a reasonable approximation where photon noise dominates. Finally, we highlight that whether a planet in the CHZ is truly habitable is a completely different question that we make no attempt to investigate in this work.

Ultimately, our work makes a strong case that although LSST may not have been built with WD transits as a science case in mind, it is uniquely placed to conduct the most in-depth survey to date. A preliminary survey with \textit{K2} of 1148 WDs by \citet{K2:2018} found no transiting objects and placed upper limits on planet occurrence rates in the range of $\eta=25\%$ to $\eta=95\%$ for $4$\,h to 10\,d roughly Earth-sized bodies. For comparison, if LSST failed to detect similar objects, the occurrence rate would be constrained to be $\eta \lesssim 0.05$\%. With prior indications of planetary material falling onto $\sim 30$\% of WDs \citep{zuckerman:2003,zuckerman:2010,koester:2014}, a search for minor bodies around such stars is both timely and critical for advancing our understanding of these intriguing environments.